\documentclass{aa}
\usepackage[dvips]{graphics}
\usepackage{latexsym}
\def\gtsima{$\; \buildrel > \over \sim \;$}
\def\ltsima{$\; \buildrel < \over \sim \;$}
\def\prosima{$\; \buildrel \propto \over \sim \;$}
\def\gsim{\lower.5ex\hbox{\gtsima}}
\def\lsim{\lower.5ex\hbox{\ltsima}}
\def\simgt{\lower.5ex\hbox{\gtsima}}
\def\simlt{\lower.5ex\hbox{\ltsima}}
\def\simpr{\lower.5ex\hbox{\prosima}}

\def\h1{$h^{-1}$}
\def\beq{\begin{equation}}
\def\eeq{\end{equation}}
\begin{document}
%\thesaurus{03(???)}
\title{
The K20 survey. I. Disentangling old and dusty star-forming 
galaxies in the ERO population\thanks{Based 
on observations made at the European Southern Observatory,
Paranal, Chile (ESO LP 164.O-0560).}}
%The EIS observations have been 
%carried out using the ESO New Technology Telescope (NTT) at the La Silla 
%observatory under Program-ID Nos. 61.A-9005(A), 162.O-0917, 163.O-0740,
%164.O-0561
\author{
	A. Cimatti \inst{1}
	\and
	E. Daddi \inst{2}
	\and
	M. Mignoli \inst{3}
	\and
	L. Pozzetti \inst{3}
	\and
	A. Renzini \inst{4}
	\and
	G. Zamorani \inst{3}
	\and
	T. Broadhurst \inst{4,9}
	\and
	A. Fontana \inst{5}
	\and
	P. Saracco \inst{6}
	\and
	F. Poli \inst{7}
	\and
	S. Cristiani \inst{8}
	\and
	S. D'Odorico \inst{4}
	\and
	E. Giallongo \inst{5}
	\and
	R. Gilmozzi \inst{4}
	\and
	N. Menci \inst{5}
}
\institute{ 
Osservatorio Astrofisico di Arcetri, Largo E. Fermi 5, I-50125 Firenze, Italy 
\and Dipartimento di Astronomia, Universit\`a di Firenze, Largo E. Fermi 5, 
I-50125 Firenze, Italy 
\and Osservatorio Astronomico di Bologna, via Ranzani 1, I-40127, Bologna, 
Italy
\and European Southern Observatory, Karl-Schwarzschild-Str. 2, D-85748, 
Garching, Germany
\and Osservatorio Astronomico di Roma, via Dell'Osservatorio 2, Monteporzio, 
Italy
\and Osservatorio Astronomico di Brera, via E. Bianchi 46, Merate, Italy
\and Dipartimento di Astronomia, Universit\`a ``La Sapienza'', Roma, Italy
\and ST, European Coordinating Facility, Karl-Schwarzschild-Str. 2, D-85748, 
Garching, Germany
\and Racah Institute for Physics, The Hebrew University, Jerusalem, 91904, Israel
}
\offprints{Andrea Cimatti, \email{cimatti@arcetri.astro.it}}
\date{Received ; accepted }
\abstract{We present the results of VLT optical spectroscopy of a 
complete sample of 78 EROs with $R-Ks\geq5$ over a field of 52
arcmin$^2$. About 70\% of the 45 EROs with 
$Ks\leq19.2$ have been spectroscopically identified with old
passively evolving and dusty star-forming galaxies at $0.7<z<1.5$. 
The two classes are about equally populated and
for each of them we present and discuss the average spectrum.
From the old ERO average spectrum and for $Z=Z_{\odot}$ we derive a
minimum age of $\sim$3 Gyr, corresponding to a formation redshift
of $z_f$\gtsima 2.4. PLE models with such formation redshifts
well reproduce the density of old EROs (consistent with being
passively evolving ellipticals), whereas the predictions of 
the current hierarchical merging models are lower than the 
observed densities by large factors (up to an order of magnitude).
From the average spectrum of the star-forming EROs we estimate a 
substantial dust extinction with $E(B-V)$\gtsima 0.5. The 
star formation rates, corrected for the average reddening, suggest 
a significant contribution from EROs to the cosmic star-formation 
density at z$\sim$1. 
} 
\titlerunning{EROs in the K20 survey}
\authorrunning{A. Cimatti et al.}  \maketitle

\section{Introduction}

Extremely red objects (EROs, here defined with $R-Ks>5$) were 
discovered serendipitously a decade ago (Elston et al. 1988), and recent 
wide-field surveys revealed that they form a substantial 
population (Thompson et al. 1999; Daddi et al. 2000, D00 
hereafter; McCarthy et al. 2001). Having the colors expected 
for high-$z$ old and passively evolving galaxies, EROs offer 
the opportunity to test whether the present-day massive
ellipticals formed at early cosmological times ($z_f>$2--3) with
a subsequent passive and pure luminosity evolution (PLE), or whether
they formed more recently through the merging of spiral galaxies 
(e.g. Baugh et al. 1996; Kauffmann 1996).
Studies on small fields claimed a deficit of EROs, thus favouring the 
hierarchical merging scenario (e.g. Zepf 1997; Barger et al. 1999; 
Rodighiero et al. 2001), but recent surveys on wider fields showed 
that the surface density of EROs is consistent with elliptical galaxy
PLE expectations (D00; Daddi, Cimatti \& Renzini 2000). Since only a 
few old galaxies have been spectroscopically identified (e.g. 
Spinrad et al. 1997; Cohen et al. 1999), their fraction among EROs
remained still unconstrained. 
On the other hand, EROs may also be high-$z$ starbursts and AGNs strongly 
reddened by dust extinction. Such a possibility was confirmed by 
the identification of this kind of galaxies among EROs, but, again, 
these results were limited to a handful of objects (e.g. Graham \& Dey
1996; Cimatti et al. 1998; Gear et al. 2000; Pierre et al. 2001, 
Smith et al. 2001, Afonso et al. 2001), thus leaving undetermined the 
relative fractions of old and dusty galaxies in ERO samples.

Because of the stringent test of galaxy formation scenarios
that EROs can provide, it is therefore of prime importance
to determine the relative fractions among the two classes of galaxies. 
In this Letter, 
we report on the first results of deep VLT optical spectroscopy 
of a complete and sizeable sample of EROs.
A cosmology with $H_0=70$ km s$^{-1}$ Mpc$^{-1}$, 
$\Omega_m=0.3$ and $\Omega_{\Lambda}=0.7$ is adopted.

\section{The K20 ERO sample} 

The selection and the observations of the ERO sample were made in the
context of the K20 survey ({\tt http://www.arcetri.astro.it/$\sim$k20/}). 
The prime aim of such a survey is to derive the redshift distribution of 
about 550 $K$-selected objects with $Ks\leq20$ in order to constrain 
the galaxy formation models. The targets were selected from a 32.2 
arcmin$^2$ area of the Chandra Deep Field South (CDFS; Giacconi et al. 
2001) using the images from the ESO Imaging Survey 
public database (EIS; {\tt http://www.eso.org/science/eis/}; the 
$R$- and the $Ks$-band images were reduced and calibrated by the 
EIS team and by our group respectively), and
from a 19.8 arcmin$^{2}$ field centered at 0055-269 using NTT+SOFI
and VLT-UT2+FORS2 $Ks$- and $R$-band images respectively (Fontana et 
al. in preparation). More details on the photometry will be given in 
forthcoming papers. From the total sample with $Ks\leq20.0$, we extracted 
the subsample of EROs with $R-Ks\geq5.0$, with the colors
measured in 2$^{\prime\prime}$ arcsec diameter corrected aperture 
to match the color definition of D00. The total sample 
includes 78 EROs, corresponding to a surface density of 1.50$\pm$0.17 
arcmin$^{-2}$, consistent with that of Thompson et al. (1999) (1.88$\pm$
0.11 arcmin$^{-2}$ over a field of 154 arcmin$^{2}$). 
The ratio between the number of EROs and the total number of objects at 
$Ks\leq19.2$ is 0.134$\pm$0.021, consistent with the value of 
0.127$\pm$0.006 of D00. 

Multi-object spectroscopy of EROs was made with the ESO VLT UT1 and UT2 
equipped with FORS1 (October-November 1999) and FORS2 (November 2000) 
during 0.5$^{\prime\prime}$-1.5$^{\prime\prime}$ seeing 
conditions and with 0.7$^{\prime\prime}$-1.2$^{\prime\prime}$ wide 
slits depending on the seeing. The grisms 150I, 200I, 300I were used
with typical integration times of 1-3 hours. Dithering of 
the targets along the slits between two fixed positions was made for 
most observations in order to efficiently remove the CCD fringing and the 
strong OH sky lines at $\lambda_{obs}>7000$~\AA. The spectra were 
calibrated using standard spectrophotometric stars, dereddened for 
atmospheric extinction, corrected for telluric absorptions and 
scaled to the total $R$-band magnitudes. 

\begin{table}
\caption[]{The ERO spectroscopic identifications}
\begin{tabular}{llllllc} \hline
Sample & $N_{tot}$ & $N_{id}$ & $N_{u}$ & $N_{old}$ & $N_{sf}$ & $N(ambig)$ \\
 & & & & & & \\
$Ks \leq 20.0$ & 78 & 35 & 43 & 15 & 18 & 2 \\
$Ks \leq 19.2$ & 45 & 30 & 15 & 14 & 15 & 1 \\
\hline
\end{tabular}
\end{table}

\section{The ERO spectral types and relative fractions}

The spectroscopic analysis was done by means of automatic software (IRAF: 
{\tt rvidlines} and {\tt xcsao}) and through visual inspection of the 
1D and 2D spectra. The redshift distributions are shown in Fig. 1.

According to the observed features, EROs were grouped into two classes:
{\it (1) old} if the spectrum showed no emission lines, a prominent
D4000 break, strong CaII H\&K absorptions, and the overall shape
of the continuum expected in case of an {\it old passively evolving 
elliptical},
{\it (2) star-forming} when the spectrum showed a clear [OII]$\lambda$3727
emission and no evidence for a strong D4000 break. Among the 15 old
EROs, four present weak [OII]$\lambda$3727 emission. We also identified two 
galaxies at $z>1.6$ thanks to MgII$\lambda$2800 and FeII$\lambda$2600 
absorptions, but their nature remained ambiguous because of the lack of other
distinctive spectral features.
Table 1 summarizes the results, giving the numbers of spectroscopically 
identified and unidentified EROs ($N_{id}$ and $N_u$), of old ($N_{old}$) 
and star-forming galaxies ($N_{sf}$). 

\begin{figure}[ht]
\resizebox{\hsize}{!}{\includegraphics{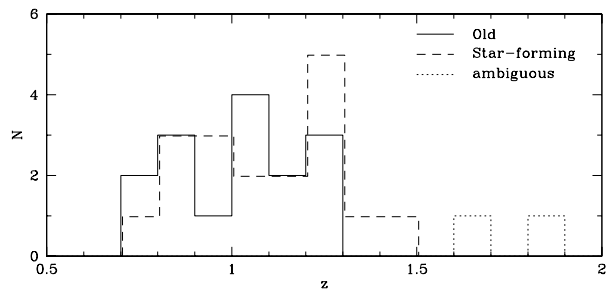}}
\caption{\footnotesize 
The redshift distributions of the identified EROs (the distribution
of star-forming EROs is slighlty shifted in $z$ to improve the 
visibility of its difference from that of old EROs). 
Due to the 
noise at $\lambda_{obs}>$9000~\AA, our spectroscopy makes it 
difficult to identify old and star-forming galaxies 
at $z>1.3$ and $z>1.5$ respectively. The redshift distribution 
of the old EROs is broadly consistent with those discussed 
by Daddi et al. (2001), McCarthy et al. (2001) and Firth et al.
(2001).
}
\label{fig:plot}
\end{figure}

In this paper, we limited the analysis to the $Ks\leq19.2$ 
subsample of EROs because it provides the best compromise between 
spectroscopic completeness (67\% of the EROs identified vs. 44\% at
$Ks\leq 20$) and significant statistics. In such a subsample, 
the fractions of old and star-forming galaxies are 
respectively $N_{old}/ N_{tot}= 31\% \div 64\%$ and $N_{sf}/N_{tot}=33\% 
\div 67\%$, where the range of such fractions corresponds to assuming 
that none or all of the unidentified objects belong to the group. 
Such fractions are generally consistent with the results of HST 
imaging by Moriondo et al. (2000) and Stiavelli \& Treu (2000).

\section{The ERO average spectra}

In this Letter we limit the discussion to the average spectra of 
the identified galaxy types in order to derive their main properties 
that would be otherwise difficult to characterize in the individual
noisier spectra. The average spectra shown in Fig. 2, obtained 
deredshifting (with a 5~\AA~ rest-frame bin), normalizing and 
stacking the equally weighted individual spectra, are relative 
to all spectroscopically identified EROs with $Ks\leq20.0$, but they 
do not significantly change if only the $Ks\leq19.2$ EROs are used. 

\subsection{Old EROs}

By comparing the average spectrum of old EROs with Bruzual \& Charlot 
(2000, private communication) simple stellar population (SSP) models 
(Salpeter IMF, $Z=Z_{\odot}$) and no dust extinction, and taking 
into account the observed average $R-Ks$ color (5.19$\pm$0.06), we 
derive an average age of 3.3$\pm$0.3 Gyr (Fig. 3). The amplitude of 
the average D4000 break (1.92$\pm$0.06) is consistent with an age 
of 2.8$\pm$0.3 Gyr. If we adopt an average age of 3.1$\pm$0.3 Gyr 
for $Z=Z_{\odot}$, the mean formation redshift is then $z_f=2.4 \pm 
0.3$. If an e-folding time of star formation $\tau=0.3$ Gyr is adopted, 
or if a Scalo (1986) IMF is used, the age increases to about 4 Gyr. 
SSP models with a lower metallicity ($Z=0.4Z_{\odot}$) would further
increase the age to $\sim 5-6$ Gyr leading to extremely high $z_f$. 
On the other hand, a higher metallicity with $Z=2.5Z_{\odot}$ would 
reduce the age to $\sim$1.1 Gyr and the formation redshift to $z_f\sim 
1.5$, but it would underestimate the observed $R-Ks$ colors of old 
EROs at $z>1$, thus being an inappropriate possibility for the
highest redshift objects.

To summarize, being SSP models with an instantaneous burst of
star formation rather unrealistic, the age of $\sim 3$ Gyr and
$z_f=2.4$ should be considered lower limits if $Z=Z_{\odot}$. 

\subsection{Star-forming EROs}

As a first attempt to investigate the nature of the star-forming 
EROs, we compared the global shape of their average spectrum 
with template spectra of star-forming galaxies, although the
presence of [NeV]$\lambda$3426 emission with an equivalent 
width $W=3.6 \pm 0.7$~\AA~ may indicate a more complex picture 
with also a contribution from dust-obscured AGN activity.
A more detailed 
analysis based on absorption lines with an equivalent 
width of a few ~\AA~ is not warranted by our data because of the 
limited signal to noise ratio. 
Among the Kinney et al. (1996) templates, a good agreement is 
found only with their so called SB6 spectrum (i.e. the average 
spectrum of starburst galaxies with $0.6<E(B-V)<0.7$ as derived from 
the H$\alpha$/H$\beta$ ratio), but only if the reddening is increased 
by an additional $E(B-V)\sim0.5$ (we adopt the Calzetti et al. 2000 
extinction curve throughout the paper). 
Since the stellar continuum and the ionized gas of dusty starbursts
suffer different extinctions ($E(B-V)_{star}\sim 0.44E(B-V)_{gas}$;
Calzetti et al. 2000), the net total extinction of the continuum
of the ERO average spectrum is $E(B-V)_{star}\sim (0.65 \times 0.44)+
0.5 \sim 0.8$.

The average spectrum of e(a) VLIGs (Very Luminous Infrared Galaxies;
${\rm log}(L_{IR}/L_{\odot})>11.5$; Poggianti \& Wu 2000) also provides 
a global satisfactory agreement at $\lambda>3600$~\AA~ without the need 
of extra dust extinction (the median reddening estimated for the e(a) 
galaxies is $E(B-V)_{gas}$=1.1 based on H$\alpha$/H$\beta$ ratio, 
corresponding to $E(B-V)_{star}\sim 0.5$; Poggianti \& Wu 2000). 

Finally, a comparison with synthetic spectra of star-forming galaxies 
with solar metallicity, Salpeter IMF and constant star formation rate 
(SB99 models, Leitherer et al. 1999 and Bruzual \& Charlot 2000 models) 
showed that the global shape of the continuum and the average $R-Ks$ 
color can be reproduced with a wide range of ages and with $0.6<E(B-V)<1.1$.

The possibility that a fraction of star-forming EROs have an old 
bulge component contributing to the red colors (e.g. similar to 
the red massive disk galaxy at $z=1.34$ of van Dokkum \& Stanford 
2001) is not ruled out by our data. In fact, HST imaging already  
showed that the "non-elliptical" EROs are morphologically made by 
a heterogeneous population ranging from highly irregular systems 
(likely to be the most dusty starbursts) to disky galaxies 
(Moriondo et al. 2000; Stiavelli \& Treu 2000). The noise in our 
spectra hampers a detailed analysis of the individual EROs. However,
by subtracting the average spectrum of old from that of star-forming 
EROs, we estimate that at most 30-40\% of the total light of the 
star-forming EROs at $\sim$4000~\AA~ can be due to an old system. 
In such a case, the average reddening would decrease to $E(B-V)_{star}
\sim$0.7 using the SB6 template to reproduce
the ``pure'' star-forming component.

\section{The density of old EROs and limits on $z_f$}

Since old EROs have spectra consistent with being passively evolving
ellipticals, we compared their density with different model predictions.

\subsection{The surface density}

The total surface density of EROs with $Ks\leq19.2$ and $R-Ks\geq5.0$ 
in our sample is 0.88$\pm$0.13 arcmin$^{-2}$, consistent with that derived
by D00 and Firth et al. (2001) over larger fields. 
In marked contrast, the surface density of EROs (counted as old 
plus dusty galaxies) predicted by the 
hierarchical merging models at $K=19.2$ presented by Firth et al. 
(2001) and Smith et al. (2001) are below the observed density by 
factors of $\sim$4 and about an order of magnitude respectively.

From the minimum and maximum fractions of old EROs derived in 
Section 3, the implied surface densities of ellipticals at the same 
magnitude limit should be in the range 0.27-0.55 arcmin$^{-2}$.
Within the uncertainties, such densities agree with those predicted 
by the PLE models of Daddi et al. (2000b) with $\tau$=0.3 Gyr
(0.24 and 0.54 arcmin$^{-2}$ for $z_f$=2.2 and $z_f$=3.5 respectively),
suggesting a {\it minimum} formation redshift of $z_f\sim 2.2$,
consistent with the age derived from the average spectrum.
The result does not significantly change if the 4 old EROs with
weak [OII]$\lambda$3727 emission are excluded from the sample. 
If $\tau$=0.1 Gyr is adopted, or if the minimum fraction of old EROs 
(31\%) is applied to the 
surface density of EROs derived by D00 in a much wider field 
(0.67 arcmin$^{-2}$ over 450 arcmin$^{2}$), or if the 2MASS $K$-band 
local luminosity function of early type galaxies is used (Kochanek 
et al. 2001), the {\it lower limit} becomes $z_f\sim 2$.

\subsection{The comoving density}

Taking into account the nominal lower redshift limit imposed by the 
$R-Ks\geq5.0$ color cut and the upper boundary due to the
sensitivity of our spectroscopy (0.85\ltsima $z$\ltsima 1.30), 
we derive a comoving density of ellipticals in that redshift
range of (2.16$\pm$0.62) $\times 10^{-4}$ Mpc$^{-3}$.
Such a density can be considered a lower limit because some 
ellipticals can escape the threshold of $R-Ks\geq5.0$ due to the 
dispersion of the observed $R-Ks$ colors around the model color 
prediction. Even so, the densities expected in the Daddi et al. (2000b) 
PLE model with $\tau$=0.3 Gyr are in the range of (1.8,2.7) 
$\times 10^{-4}$ Mpc$^{-3}$ for $z_f\sim$(2.2,2.4), thus being 
consistent with the observed densities.

\section{The role of dusty star-forming EROs}

The observed star formation rate (SFR) of each 
star-forming ERO has been estimated both from the [OII]$\lambda$3727 
line emission and the 2800~\AA~ continuum luminosities using the 
relations of Kennicutt (1998). 
The average SFRs at $z_{mean}=1.096$ are SFR([OII])=3.6 M$_{\odot}$
yr$^{-1}$ and SFR(L$_{2800}$)= 1.8 M$_{\odot}$ yr$^{-1}$. 
Adopting the comoving volume included between the observed $z_{min}$ and 
$z_{max}$, the corresponding SFR densities are SFRD([OII])=0.0011 
and SFRD(L$_{2800}$)=0.0005 M$_{\odot}$ yr$^{-1}$ Mpc$^{-3}$. Such 
SFRDs clearly represent {\it lower limits} because no corrections 
for dust extinction and for incompleteness have been applied. 
However, such SFRs may also be partly overestimated if dust-obscured 
AGN activity is present in some EROs as suggested by the possible 
[NeV]$\lambda$3426 emission (Fig.1).

If we conservatively adopt an average $E(B-V)\sim 0.5$ (see Section 4)
and apply the corresponding extinction corrections, the ERO SFRD becomes
$\sim 0.015$ M$_{\odot}$ yr$^{-1}$ Mpc$^{-3}$, formally corresponding 
to a contribution of about 20\% to the global SFRD of the universe at 
$z\sim 1$ (without counting EROs) corrected for dust extinction 
($\sim 0.08$ M$_{\odot}$ yr$^{-1}$ Mpc$^{-3}$, as discussed by 
Somerville et al. 2001). Even if the uncertainties are large because 
of the assumptions on the SFR estimators and on the adopted extinction 
curve, our result strongly suggests that EROs may be important 
in the cosmic star formation budget at $z\sim 1$. 

Our results also suggest that the ERO selection provides the possibility 
to uncover the population of high-$z$ dusty star-forming galaxies in 
a way complementary to the surveys for submillimeter/millimeter-selected
galaxies. In fact, if the dereddened SFRs of star-forming EROs are in the
range of 50-150 M$_{\odot}$ yr$^{-1}$ (possible for $E(B-V)\sim
0.5-0.7$), this would suggest that their far-infrared luminosities 
are generally below $10^{12}$ L$_{\odot}$ (adopting the relationship
SFR[M$_{\odot}$ yr$^{-1}$]=$4.5\times10^{-44}$L$_{FIR}$[erg s$^{-1}$];
Kennicutt 1998). Such a scenario would explain the origin of the 
low detection rates of EROs in submillimeter/millimeter continuum 
follow-up observations, typically sensitive to detect ultra-luminous 
infrared galaxies (ULIGs, L$>10^{12}$ L$_{\odot}$) at $z$\gtsima 1 
(e.g. Mohan et al. 2001).

\begin{figure}[ht]
\resizebox{\hsize}{!}{\includegraphics{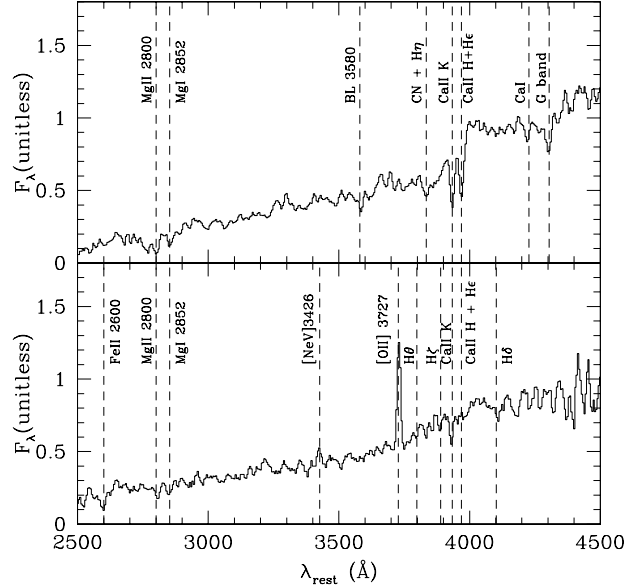}}
\caption{\footnotesize 
The average rest-frame spectra (smoothed with a 3 pixel boxcar) of 
old passively evolving (top; $z_{mean}=1.000$) and dusty 
star-forming EROs (bottom; $z_{mean}=1.096$) with $Ks\leq20$.
}
\label{fig:plot}
\end{figure}

\begin{figure}[ht]
\resizebox{\hsize}{!}{\includegraphics{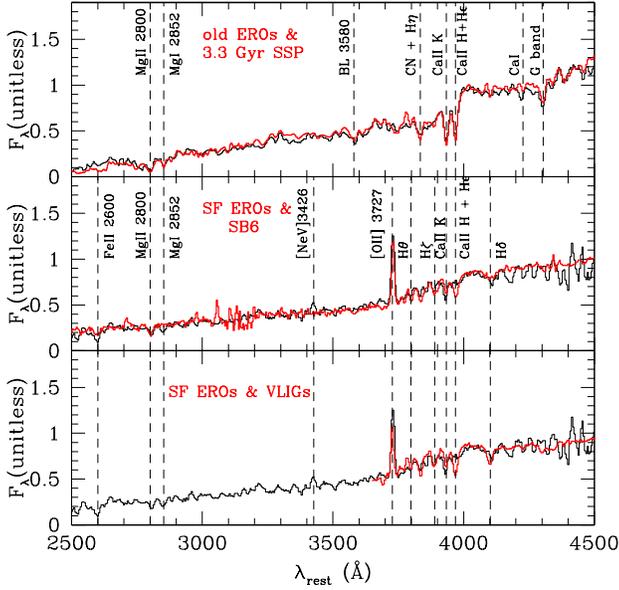}}
\caption{\footnotesize 
The average spectra of EROs (thin lines) and template galaxies 
(thick lines) (see text for details). The spikes at 
$\lambda\approx$3100~\AA~ are due to the noise in the SB6 spectrum.
}
\label{fig:plot}
\end{figure}

\begin{acknowledgements}
We thank the referee, P. McCarthy, for the constructive comments,
and the VLT support astronomers for their kind assistance.
AC warmly thanks ESO (Garching) for the hospitality during
his visits, A. Franceschini for useful discussion and B. Poggianti 
for providing the VLIG spectrum. 
\end{acknowledgements}

\end{document}